\documentclass[fleqn,10pt]{wlscirep}
\usepackage[utf8]{inputenc}
\usepackage[T1]{fontenc}
\usepackage{graphicx}
\newcommand*{\oi}{{\rm i}}
\title{Dynamical Quantum Phase Transition and Quasi Particle Excitation}

\author[1,2,3*]{R. Jafari}
\affil[1]{Department of Physics, Institute for Advanced Studies in Basic Sciences (IASBS), Zanjan 45137-66731, Iran}
\affil[2]{Beijing Computational Science Research Center, Beijing 100094, China}
\affil[3]{Department of Physics, University of Gothenburg, SE 412 96 Gothenburg, Sweden}
\affil[*]{jafari@iasbs.ac.ir, rohollah.jafari@gmail.com}

\begin{abstract}
Dynamical phase transitions (DPTs) are signaled by the non-analytical time evolution of the dynamical free energy
after quenching some global parameters in quantum systems. The dynamical free energy is calculated from the
overlap between the initial and the time evolved states (Loschmidt amplitude). In a recent study it was suggested
that DPTs are related to the equilibrium phase transitions (EPTs) (M. Heyl et al., Phys. Rev. Lett. \textbf{110}, 135704 (2013)).
We here study an exactly solvable model, the extended $XY$ model, the Loschmidt amplitude of which provides a counterexample.
We show analytically that the connection between the DPTs and the EPTs does not hold generally. Analysing also the general
compass model as a second example, assists us to propound the physical condition under which the DPT occurs without crossing
the equilibrium critical point, and also no DPT by crossing the equilibrium critical point.
\end{abstract}
\begin{document}

\flushbottom
\maketitle
%
%
\thispagestyle{empty}


\section*{Introduction}

Recently, the study of non-equilibrium properties of quantum systems have been
attracting a lot of attention \cite{Montes, Happola, Zurek2005, Kennes, Quan}.
One of the ongoing interest is to understand the notion of universality for a system away from equilibrium.
Recent advancement in the studies of ultra-cold atoms trapped in optical lattices provide a novel framework
to prob the non-equilibrium dynamics of quantum critical phenomena \cite{Bloch, Chen2011, Chen2014, Polkovnikov2011}.
Specifically, by considering a quantum quench, where a system is prepared in a well defined initial state
and then suddenly changing the external parameters in the Hamiltonian controls the unitary evolution of the system \cite{Dutta, Kolodrubetz2012, Campbell}.
The non-equilibrium dynamics of the quenched quantum system can be probed in many different ways, borrowing ideas from equilibrium
statistical mechanics. In a recent work the notion of dynamical phase transitions (DPTs) has been introduced
probing the non-analyticities in the dynamical free energy in the complex time plane \cite{Heyl2013}.
The idea originates from the resemblance between the canonical partition function
of an equilibrium system $Z(\beta)=Tr e^{-\beta{\cal H}}$ and that of the quantum boundary partition function
$Z(z)=\langle\psi_{0}|e^{-z{\cal H}}|\psi_{0}\rangle$ \cite{LeClair, Piroli} which corresponds to the Loschmidt amplitude (LA)
for $z=it$. The LA ($L(t)=\langle\psi_{0}(h^{(1)})|e^{-i{\cal H}(h^{(2)})t}|\psi_{0}(h^{(1)})\rangle$) is the overlap
amplitude of the initial quantum state $|\psi_{0}(h^{(1)})\rangle$ with its time evolved state under the post-quenched
Hamiltonian ${\cal H}(h^{(2)})$. In the complex time ($z$) plane, the dynamical free energy density is defined as
$f(z)=-\lim_{N\rightarrow\infty}\ln Z(z)/N$  where $N$ is the number of degrees of freedom \cite{Heyl2013, Vajna, Andraschko, Divakaran2016}.
In a spirit similar to the classical case, one then looks for the non-analyticities of $f(z)$ or zeros of the $Z(z)$,
known as Fisher zeros where interpreted as a dynamical phase transition \cite{Heyl2013, Vajna, Andraschko, Heylrev}.
Additionally, these DPTs are presented in sharp nonanalyticities in the rate function of the return probability
(Loschmidt echo) defined as $l(t)=-\lim_{N\rightarrow\infty}\ln |L(t)|^{2}/N$ \cite{Pollmann, Heyl2013, Andraschko, Sharma2015,Jada,Jadb,Jadc,Jadd,Jade}.

A similar observation was first made by M. E. Fisher \cite{Fisher}, who pointed out that the phase transition in a thermodynamic
system is signaled by the non-analyticities in the free-energy density of an equilibrium system whose information can be acquired
by analyzing the zeros of the partition function in a complex temperature plane. These zeros of the partition
function cutting the real axis in the thermodynamic limit and integrate into a line in complex temperature plane \cite{Saarloos}.
These crossings mark the non-analyticities in the free-energy density. A similar observation was reported earlier for a complex magnetic plane
by Lee-Yang \cite{Yang}.

An initial analytical result for the dynamical phase transition in the one-dimensional transverse Ising model \cite{Heyl2013}
was verified in several subsequent studies for both integrable \cite{Heylrev, Zvyaginrev} and non-integrable models \cite{Vajna, Karrasch, Kriel, Canovi, Palmai, Andraschko, Sharma2015, Heylrev, Zvyaginrev}
which established that the DPTs occur only if the sudden quench crosses the equilibrium quantum critical point.
These works have been extended to the higher dimensional systems \cite{Vajna2015, Schmitt}, the dynamical topological order parameter \cite{Budich},
the role of topology \cite{Vajna2015}, and slow quench scopes \cite{Sharma2016, Divakaran2016}.
Further studies, however, reveal that DPTs can occur following a sudden quench even within the same phase
(i.e., not crossing the QCP) for both non-integrable \cite{Sharma2015, Andraschko, Heylrev} as well as integrable models \cite{Vajna}.
This distinct property can be emanated from a kinetic constraint. The kinetic constraint is a $U(1)$ symmetry due to magnetization (particle)
conservation which does not allow to dynamically enter the magnetization sectors (particle number) where the system adopts in the equilibrium case \cite{Andraschko ,Heylrev}.

To the best of our knowledge, there has been no general principle to connect the DPTs to the QPTs.
The purpose of this paper is to highlights the physical conditions under which the
quantum system may show DPT. To this aim, we serve two models as examples, the extended $XY$ chain in a staggered magnetic field and the
general compass model, to show that generally DPTs can occur in quenches crossing the point where the quasiparticles are massless.
Such quasiparticles may indeed be expected to appear at the quantum phase transition point, but as our case studies of the extended XY model
and extended quantum compass chain (EQCC) reveal, this is not necessarily so.
%
%

\section*{The extended XY model}
The extended $XY$ model dictated by the following Hamiltonian
%
%
\begin{eqnarray}
\label{eq1}
{\cal H}=-\frac{1}{2} \sum_{n=1}^N \Big[\frac{J}{2}(\sigma_n^{x}\sigma_{n+1}^{x}
+\sigma_{n}^{y}\sigma_{n+1}^{y})+\frac{J_{3}}{4}(\sigma_{n}^{x}\sigma_{n+2}^{x}+\sigma_{n}^{y}\sigma_{n+2}^{y})\sigma_{n+1}^{z}+(-1)^{n}h_{s}\sigma_{n}^{z}\Big],
\end{eqnarray}
%
%
where, $N$ is the system size, $h_{s}$ represents the staggered transverse field, $J$ and $J_{3}$ are
exchange couplings between the spins on the nearest-neighbor and the next-nearest-neighbor sites respectively.
Performing the Jordan-Wigner fermionization and introducing the Nambu spinor $\Gamma^{\dagger}=(c^{q\dagger}_{k},c^{p\dagger}_{k})$,
the Fourier transformed Hamiltonian can be expressed in
Bogoliubov-de Gennes (BdG) form \cite{Zhu2016,Titvinidze}, $H= -\sum_{k\ge0}\Gamma^{\dagger}H(k)\Gamma$, with
\begin{eqnarray}
\label{eq2}
H(k)=
\left(
  \begin{array}{cc}
    \frac{J_{3}}{2}\cos(k)+h_{s} & -J\cos(k/2)  \\
    \\
    -J\cos(k/2) &  \frac{J_{3}}{2}\cos(k)-h_{s} \\
  \end{array}
\right),
\end{eqnarray}
%
where $k=4\pi n/N$ with $-N/4<n<N/4$ for periodic boundary conditions \cite{Titvinidze}.
Using the standard Bogoliubov transformation
%
\begin{eqnarray}
\nonumber
c_{k}^{q}=\cos(\frac{\theta_{k}(h_{s})}{2}) \alpha_{k}+\sin(\frac{\theta_{k}(h_{s})}{2}) \beta_{k},~~
c_{k}^{p}=-\sin(\frac{\theta_{k}(h_{s})}{2}) \alpha_{k}+ \cos(\frac{\theta_{k}(h_{s})}{2}) \beta_{k},
\end{eqnarray}
%
where
%
\begin{eqnarray}
\label{eq3}
\tan(\theta_{k}(h_{s}))\!=\!-J\cos(k/2)/h_{s},
\end{eqnarray}
%
we finally can write the Hamiltonian in the diagonalized form as
${\cal H}\!=\!\sum_{k}[\varepsilon^{\alpha}_{k}(h_{s})\alpha^{\dagger}_{k} \alpha_{k}
+\varepsilon^{\beta}_{k}(h_{s})\beta^{\dagger}_{k}\beta_{k}]$,
where
%
\begin{eqnarray}
\nonumber
\varepsilon^{\alpha}_{k}(h_{s})\!=\!(J_{3}/2)\cos(k)-\sqrt{(h_{s})^{2}+J^{2}\cos^{2}(k/2)},~~
\varepsilon^{\beta}_{k}(h_{s})\!=\!(J_{3}/2)\cos(k)+\sqrt{(h_{s})^{2}+J^{2}\cos^{2}(k/2)},
\end{eqnarray}
%
%
\begin{figure*}[t]
\centerline{\includegraphics[width=\linewidth,height=0.4\linewidth]{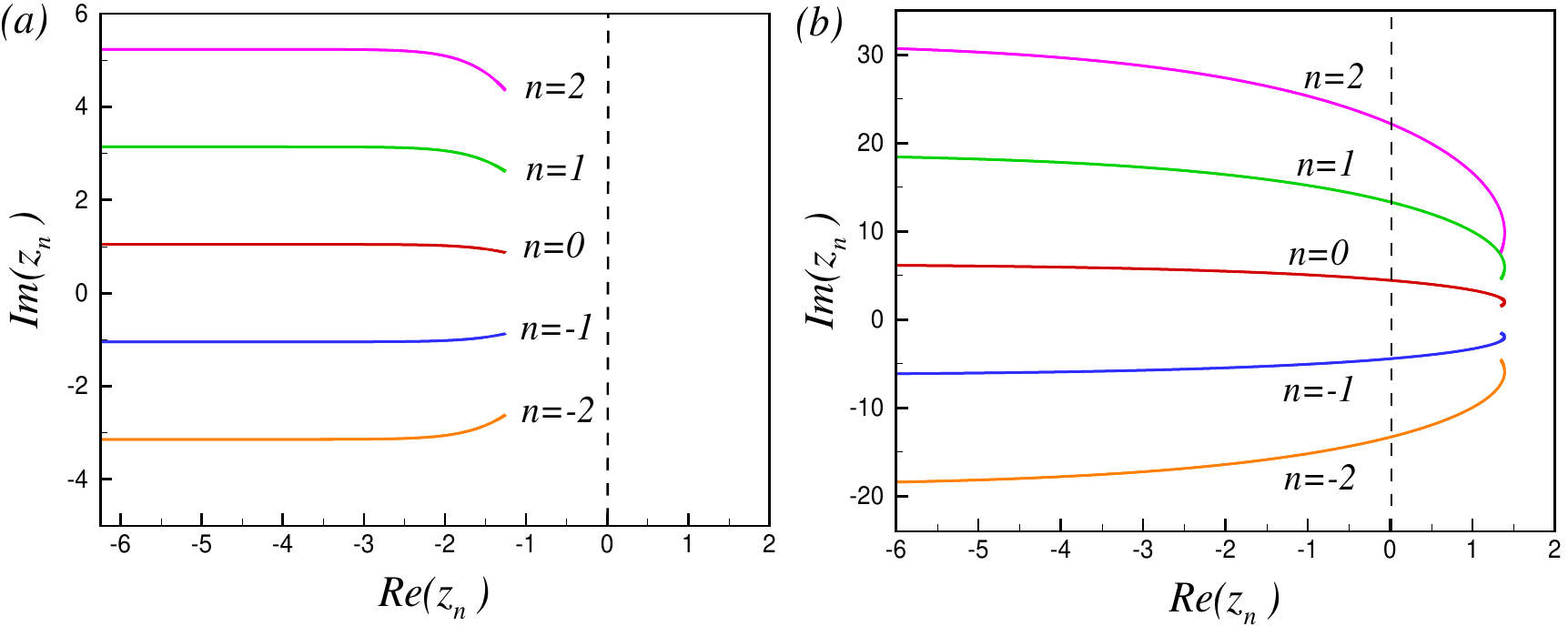}}
\caption{ (Color online) (a) Lines of Fisher zeros in the complex plane
($z$) for a quench across the gapped to gapless quantum phase transition
point at $h^{c}_{s}=2$, ($h^{(1)}_{s}=2.5$, $h^{(2)}_{s}=1.5$).
(b) Lines of Fisher zeroes for a quench within the same
phase (Spin Liquid) crossing the non-critical point $h_{c}=0$
($h^{(1)}_{s}=0.25$, $h^{(2)}_{s}=-0.25$).}
\label{fig1}
\end{figure*}
%
with corresponding quasiparticle eigenstates
%
\begin{eqnarray}
\label{eq4}
\alpha_{k}^{\dagger}|0\rangle_{k}=\cos(\frac{\theta_{k}(h_{s})}{2})c_{k}^{q\dagger}|0\rangle_{k}
-\sin(\frac{\theta_{k}(h_{s})}{2})c_{k}^{p\dagger}|0\rangle_{k},~~
\beta_{k}^{\dagger}|0\rangle_{k}=\sin(\frac{\theta_{k}(h_{s})}{2})c_{k}^{q\dagger}|0\rangle_{k}
+\cos(\frac{\theta_{k}(h_{s})}{2})c_{k}^{p\dagger}|0\rangle_{k},
\end{eqnarray}
%
where $|0\rangle_{k}$ is vacuum states of fermions.

This model reveals three phases, long-range ordered anti-ferromagnetic phase, in addition to two different
spin liquid phases, spin liquid (I) and spin liquid (II). The phase transition between anti-ferromagnetic phase
and spin liquid (I) is the gapped to gapless phase transition which occurs at $h_{s}^{c1}=\pm J_{3}/2$ (for simplicity we take $J=1$).
The system is the antiferromagnet for $|h_{s}|\geq J_{3}/2$ where $\varepsilon^{\alpha}_{k}(h_{s})\leqslant0$
and $\varepsilon^{\beta}_{k}(h_{s})>0$ for all $k$ mode, and therefore the ground state for each mode is
$\alpha_{k}^{\dagger}|0\rangle_{k}$ with the total ground state energy $E_{g}=\sum_{k}\varepsilon^{\alpha}_{k}(h_{s})$.
For $\sqrt{J_{3}^{2}/4-1}<|h_{s}|<J_{3}/2$ system enters into the spin liquid (I) phase where
$\varepsilon^{\alpha}_{k}(h_{s})\leqslant0$ for all modes in addition to $\varepsilon^{\beta}_{k}(h_{s})$
which is negative for some of the $k$ mode. So, for a given mode where both $\varepsilon^{\alpha}_{k}(h_{s})$
and $\varepsilon^{\beta}_{k}(h_{s})$ are negative the ground state is given by $\alpha^{\dagger}_{k}\beta^{\dagger}_{k}|0\rangle_{k}$
whereas for a mode where only $\varepsilon^{\alpha}_{k}(h_{s})$ is negative, $\alpha^{\dagger}_{k}|V\rangle_{k}$ is the
ground state of the system. The gapless-gapless phase transition takes place between spin liquid (I) and
spin liquid (II) at $h_{s}^{c2}=\pm\sqrt{J_{3}^{2}/4-1}$ where the topology of the Fermi surface changes \cite{Titvinidze}.
In the spin liquid (II) phase ($|h_{s}|\leq\sqrt{J_{3}^{2}/4-1}$) both $\varepsilon^{\alpha}_{k}(0)$
and $\varepsilon^{\beta}_{k}(h_{s})$ have both positive and negative branches resulting to four Fermi points,
two from each branch. Consequently, there are three possible ground states for a given $k$ mode depending on
the sign of the energies $\varepsilon^{\alpha,\beta}_{k}(h_s)$ given by $|0\rangle_{k}$, $\alpha^{\dagger}_{k}|0\rangle_{k}$,
and $\alpha^{\dagger}_{k}\beta^{\dagger}_{k}|0\rangle_{k}$ and the ground state energy is the sum over all the modes with
negative energies of each branch.
In what follows we will assume the system is prepared in the ground state of Hamiltonian Eq.\ref{eq1} corresponding
to $h^{(1)}_{s}$. At time $t=0$, we quench the staggered field strength $h^{(1)}_{s}\longrightarrow h^{(2)}_{s}$
and we evolve the initial state according to the new Hamiltonian ${\cal H}(h^{(2)}_{s})$.

It is straightforward to show that the Loschmidt amplitude (LA) and the return probability (RP) of
the extended $XY$ model are given by
%
\begin{eqnarray}
\label{eq5}
Z(z)=\prod_{k}e^{-z\varepsilon^{\alpha}_{k}(h_{s}^{(2)})}\Big[\cos^{2}(\eta_{k})+\sin^{2}(\eta_{k})e^{-z\Delta\varepsilon_{k}(h_{s}^{(2)})}\Big],
\end{eqnarray}
%
and
%
\begin{eqnarray}
\label{eq6}
|{\cal L}(t)|^{2}=\prod_{k}{\cal L}_k(t)=\prod_{k}[1-A_{k}\sin^{2}(\frac{\Delta\varepsilon_{k}t}{2})],
\end{eqnarray}
%
respectively, where, $2\eta_{k}=\theta_{k}(h^{(1)}_{s})-\theta_{k}(h^{(2)}_{s})$, $A_{k}=\sin^{2}(2\eta_{k})$, and $\Delta\varepsilon_{k}=\varepsilon^{\beta}_{k}(h^{(2)}_{s})-\varepsilon^{\alpha}_{k}(h^{(2)}_{s})$.
%
In the thermodynamic limit the zeroes of the Loschmidt amplitude in the complex plane coalesce to a family of lines
labeled by a number $n\in\mathbb{Z}$
%
\begin{eqnarray}
\label{eq7}
z_{n}(k)=\frac{1}{\Delta\varepsilon_{k}}\Big[i\pi(2n+1)+\ln(\tan^{2}(\eta_{k}))\Big],
\end{eqnarray}
%
The sketches of lines of Fisher zeros are shown in Figs. \ref{fig1}(a) and \ref{fig1}(b) for quenching the system across
the equilibrium quantum critical point ($h_{s}=2$) and the non-critical line ($h_{s}=0$), respectively.
As seen in Fig. \ref{fig1} (a), the lines of Fisher zeroes do not cut the imaginary axis for a quench crossing
the quantum critical point. While the imaginary axis is crossed by Fisher zeros lines if the system quenched through
the non-critical line ($h_{s}=0$) in Fig. \ref{fig1}(b).

The main quantity that controls the dynamical free energy is $\tan^{2}(\eta_{k})$, which
depends on the parameters of pre-quenched and post-quenched Hamiltonian.
The Fisher zeroes lines in the complex plane cross the imaginary axis only when there is a mode $k^{\ast}$
that satisfies $\tan^{2}(\eta_{k^{\ast}})=1$ ($\eta_{k^{\ast}}=\pi/4$). Using the expression
$2\eta_{k}=\theta_{k}(h^{(1)}_{s})-\theta_{k}(h^{(2)}_{s})$ and Eq. (\ref{eq3}) this condition
can be solved for $k^{\ast}$ analytically
%
\begin{eqnarray}
\label{eq8}
\cos^{2}(\frac{k^{\ast}}{2})=-h^{(1)}_{s}h^{(2)}_{s},
\end{eqnarray}
%
%
\begin{figure}[ht]
\centerline{\includegraphics[width=\linewidth,height=0.35\linewidth]{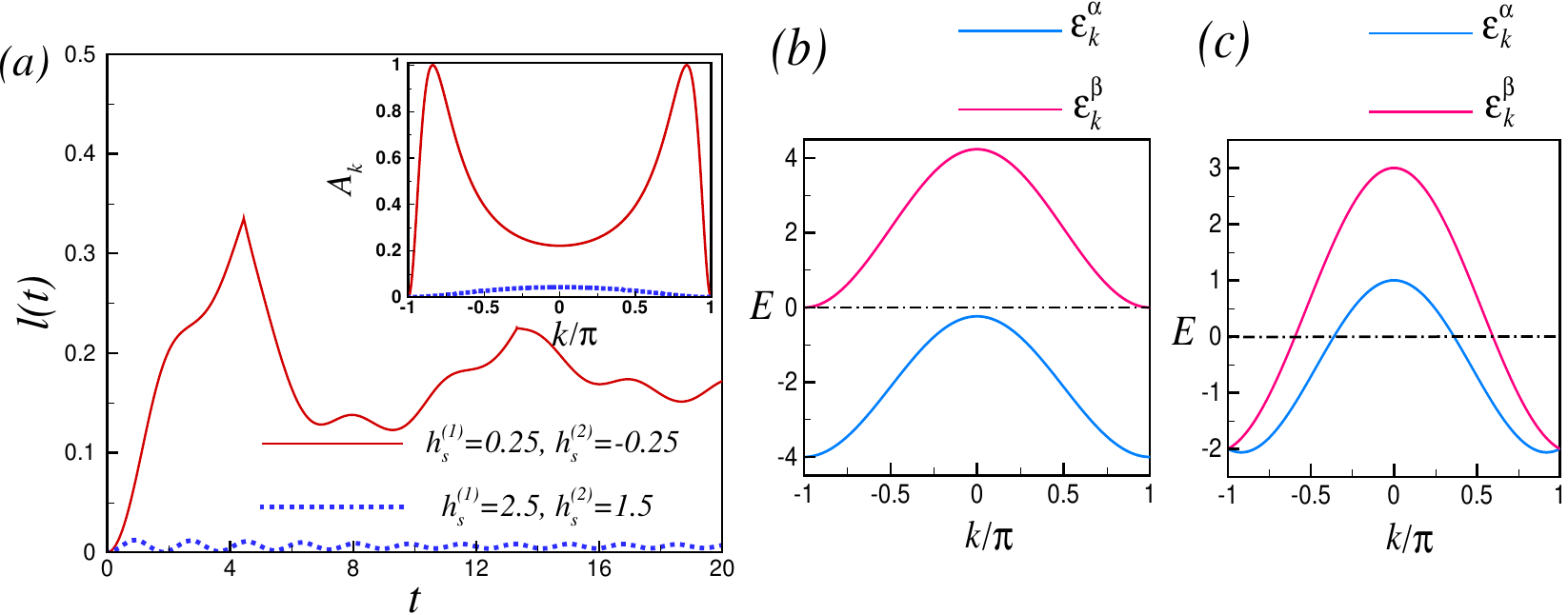}}
\caption{ (Color online) (a) The rate function of the return probability $l(t)$, for a
system with $N=120$ sites. Sharp non-analyticities in the rate function of RP
appears periodically at times $t_{n}$ (Eq. \ref{eq9}) for a quench crossing
the non-critical line $h_{s}=0$ (solid line) and wiped out for a quench
across the critical point $h^(c)_{s}=2$ (dotted line).
Inset: The amplitude of oscillation term $A_{k}$ in (Eq. \ref{eq6}) plotted
versus $k$ for a quench crossing the non-critical line
$h_{s}=0$ (solid line) and across the critical point $h^{c}_{s}=2$ (dotted line).
BdG quasiparticle spectrum ($\varepsilon^{\alpha}_{k},~\varepsilon^{\beta}_{k}$)
for the extended $XY$ model at (b) critical point $h^{c}_{s}=2$ and (c) non-critical point $h_{s}=0$.}
\label{fig2}
\end{figure}
%
It is easily seen that, the above equation can only be fulfilled provided $h^{(2)}_{s}h^{(1)}_{s}$ becomes negative.
In other words, the non-analyticities in the Loschmidt amplitude can only exist when the system is suddenly quenched through the
the non-critical line $h_{s}=0$ and $\sqrt{|h^{(2)}_{s}h^{(1)}_{s}|}\leq1$.
As a consequence of this analytical analysis the Loschmidt amplitude shows a periodic sequence of real-time
non-analyticities in the case of quenching across the non-critical line $h_{s}=0$ at times
%
\begin{eqnarray}
\label{eq9}
t_{n}=t^{\ast}(n+\frac{1}{2}),~~t^{\ast}=\frac{2\pi}{\Delta\varepsilon_{k^{\ast}}},
\end{eqnarray}
%
which numerically verified as shown in Fig. \ref{fig2}(a). We should mention that, the momentum
$k$ changes from $-\pi$ to $\pi$, and for both the quenched
cases $\lim_{k\rightarrow\pm\pi}Re[z_{n}(k)]=\pm\infty$, so each line of Fisher zeros consists of
two overlapping half lines. Then, the Fisher zeros lines in Fig. \ref{fig1}(b) cross the imaginary axis twice
which means there are two timescales in the dynamical free energy. Due to symmetric spectrum around $k=0$,
i.e., $\varepsilon_{-k}^{\alpha,\beta}=\varepsilon_{k}^{\alpha,\beta}$ two timescales are equal.

To find out why the LA shows nontrivial behaviour for the quench crossing the phase boundary
and also crossing the non-critical line let us to take a detailed look at the LE in Eq. (\ref{eq6}).
First note that, the real time instances at which the LA is zero is exactly equivalent to
the time instances at which the return probability (RP) is zero. It requires that
one factor becomes zero in Eq. (\ref{eq6}), i.e., ${\cal L}_{k^*}(t)=0$
provided that the oscillating part of a $k^{\ast}$ mode becomes one.

An analysis shows that the oscillation amplitude $A_{k}$ is small for a quench across the critical point while
it takes its maximum possible value ($A_{k}=1$) when quench performed across the line $h_{s}=0$
at $\pm k^{\ast}$ (inset Fig. \ref{fig2}(a)).
It results that the corresponding modes can contribute destructively to the RP only at $k^{\ast}$ mode at time
instances at which its oscillation term is one i.e., $A_{k^{\ast}}\sin^{2}(\Delta\varepsilon_{k^{\ast}}t/2)=1$.
Then, the equation results $t_{n}=(2n+1)\pi/\Delta\varepsilon_{k^{\ast}}=t^{\ast}(n+1/2)$ which is exactly equivalent to
Eq. (\ref{eq9}).

In other words, if the maximum possible value of the oscillation amplitude is less than $1$,
no DPT can appear even when the system is quenched across the QCPs.
It is instructive to show that the amplitude of oscillation is connected to the occupation of the excited state.
Defining excited state occupancy \cite{Kolodrubetz2012} $n_{k}=1-|\langle\psi_{0}(h^{(1)})|\psi_{0}(h^{(2)})\rangle|^{2}$ ,
where $n_{k}$ is the expectation value of the quasiparticle occupation number in the post-quench Hamiltonian
and is conserved under the time evolution, results $\sin^{2}(2\eta_{k})=4n_{k}(1-n_{k})$.
The maximum possible value of the oscillation amplitude and thus DPTs can occur whenever, $n_{k}=1/2$.
The DPT cannot occur, even the quench is performed through an equilibrium QCPs, if the value of $n_{k}$ is not equal to $1/2$,.
Since the oscillation amplitudes can be interpreted as measuring the probabilities of quasiparticle
excitations, much larger amplitudes is expected at or near the gap-closing point.
Therefore, small oscillation amplitude and consequently the absence of DPTs for a quench crossing
the antiferromagnetic-spin liquid phase transition line $h_{s}=\pm J_{3}/2$ can be attributed to the fact
that the quasiparticles which control the LA remain fully gapped as the QPT points crossed during the quench (Fig. \ref{fig2}(b)).
On the contrary, the occurrences of DPTs for a quench within the same phase (crossing the line $h_{s}=0$)
are associated with large oscillation amplitude, which is yielded from massless quasiparticle (Fig. \ref{fig2}(c)).
%
\begin{figure}[ht]
\centerline{\includegraphics[width=\linewidth,height=0.27\linewidth]{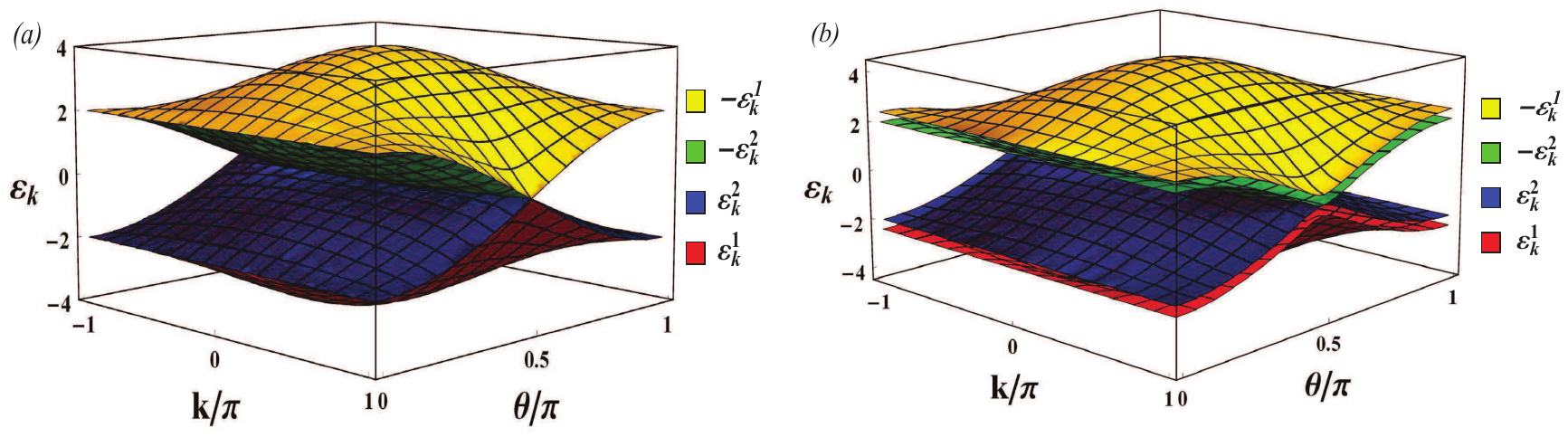}}
\caption{(Color online) Bogoliubov–de Gennes quasiparticle spectrum $\pm\varepsilon_{k}^{1,2}$
for the extended quantum compass model at (a) the isotropic point (IP) $J_{o}=J_{e}=1$, and
(b) at the anisotropic point $J_{o}\neq J_{e}$ ($J_{o}=1, J_{e}=1.2$).}
\label{fig3}
\end{figure}
%
\section*{The extended compass model}
To confirm the condition we have obtained from the study of the extended $XY$ model, under which
the dynamical phase transition occurs, we investigate the extended quantum compass model (EQCM)
as a second example.
The Hamiltonian of the spin $1/2$ extended quantum compass model (EQCM) is characterized by \cite{Henrik2017, You2014}
%
\begin{eqnarray}
\label{eq10}
{\cal H}=
\!\!
\sum_{n=1}^{N'}
\Big[
J_{o}
\tilde{\sigma}_{2n-1}^{(+)}
\tilde{\sigma}_{2n}^{(+)}
+J_{e}
\tilde{\sigma}_{2n}^{(-)}
\tilde{\sigma}_{2n+1}^{(-)}
\Big].
\end{eqnarray}
%
In this representation, on dimensional (1d) EQCM  is constructed by antiferromagnetic order of
$X$ and $Y$ pseudo-spin components on odd and even bonds at which the pseudo-spin operators are
constructed as linear combinations of the Pauli matrices $(\sigma^{\alpha=x,y,z})$:
$\tilde{\sigma}_{2n}^{(\pm)}=\tilde{\sigma}_{n}(\pm\theta)=\cos\theta\sigma^{x}_{n} \pm\sin\theta\sigma^{y}_{n}$.
Here $\theta$ ($-\theta$) is arbitrary angle relative to $\sigma^{x}$ for  even (odd) bounds.
$J_{e}$ and $J_{o}$ characterise the even and odd bound couplings respectively, and $N=2N'$ is the number of spins.
The 1d-EQCM is exactly solvable with the Jordan-Wigner transformation  \cite{Barouch},
which in  momentum space leads to ${\cal H}_E=\sum_{m=1}^{4}\sum_{k}\varepsilon^{m}_{k}\gamma_{k}^{m\dag}\gamma_{k}^{m}$,
where $\gamma_{k}^{m\dag} (\gamma_{k}^{m})$ denote independent quasiparticle creation (annihilation) operators.
For states with even fermions, $\varepsilon^{1}_{k}=-\varepsilon^{4}_{k}=\sqrt{a+\sqrt b}$ and
$\varepsilon^{2}_{k}=-\varepsilon^{3}_{k}=\sqrt{a-\sqrt b}$,
with $a_k\!=\! |J_{k}|^{2}+|L_{k}|^{2}+|J_{-k}|^{2}+|L_{-k}|^{2}$ and $b_k=4\Big[|L_{k}|^{4}+J_{k}^{2}J_{-k}^{2}-J_{k}^{\ast}J_{-k}L_{k}^{2}-J_{k}J_{-k}^{\ast}L_{-k}^{2}\Big]$,
where the parameters $L_{k}$ and $J_{k}$ are defined  by
$L_{k}=(J_o+J_e e^{\oi k})$, and $J_{k}(\theta)=(J_o e^{\oi \theta}-J_e e^{\oi (k-\theta)})$.
We concentrate on an idiosyncratic case of $\theta_c=\pi/2$ where the 1d-EQCC is critical for arbitrary
$J_{e}/J_{o}$~\cite{You2014, Nussinov2015}. QPT takes place between two different disordered phases where
the  model exhibits highest possible frustration of interactions~\cite{You2014, Nussinov2015}.

The BdG quasiparticle spectrum of the EQCC is plotted in Fig. \ref{fig3}(a)-(b) at the isotropic point (IP) $J_{o}=J_{e}$ and
at the anisotropic point $J_{o}\neq J_{e}$ respectively. The many-particle groundstate of the EQCC is obtained
by filling the two lowest bands, $\varepsilon_{k}^{1}$ and $\varepsilon_{k}^{2}$.
As seen, at the IP the energy gap between the $\varepsilon_{k}^{1}$ and $\varepsilon_{k}^{4}=-\varepsilon_{k}^{1}$ bands
closes at $\theta=\pi/2, k=\pi$ (Fig. \ref{fig3}(a)) while it is nonzero away from the IP (Fig. \ref{fig3}(b)).
In contrast, and as required for the existence of the quantum critical line $\theta_c=\pi/2$, the energy gap between
the $\varepsilon_{k}^{2}$ and $\varepsilon_{k}^{3}=-\varepsilon_{k}^{2}$ bands is closed for all $k$ at
$\theta=\pi/2$ for arbitrary values of $J_{e}/J_{o}$. One verifies that the groundstate has a
$2^{N/2}$-fold degeneracy at the critical line $\theta=\pi/2$ off the IP, with an enlarged degeneracy
$2\times 2^{N/2}$ right at the IP.
%
\begin{figure}[ht]
\centerline{\includegraphics[width=\linewidth,height=0.4\linewidth]{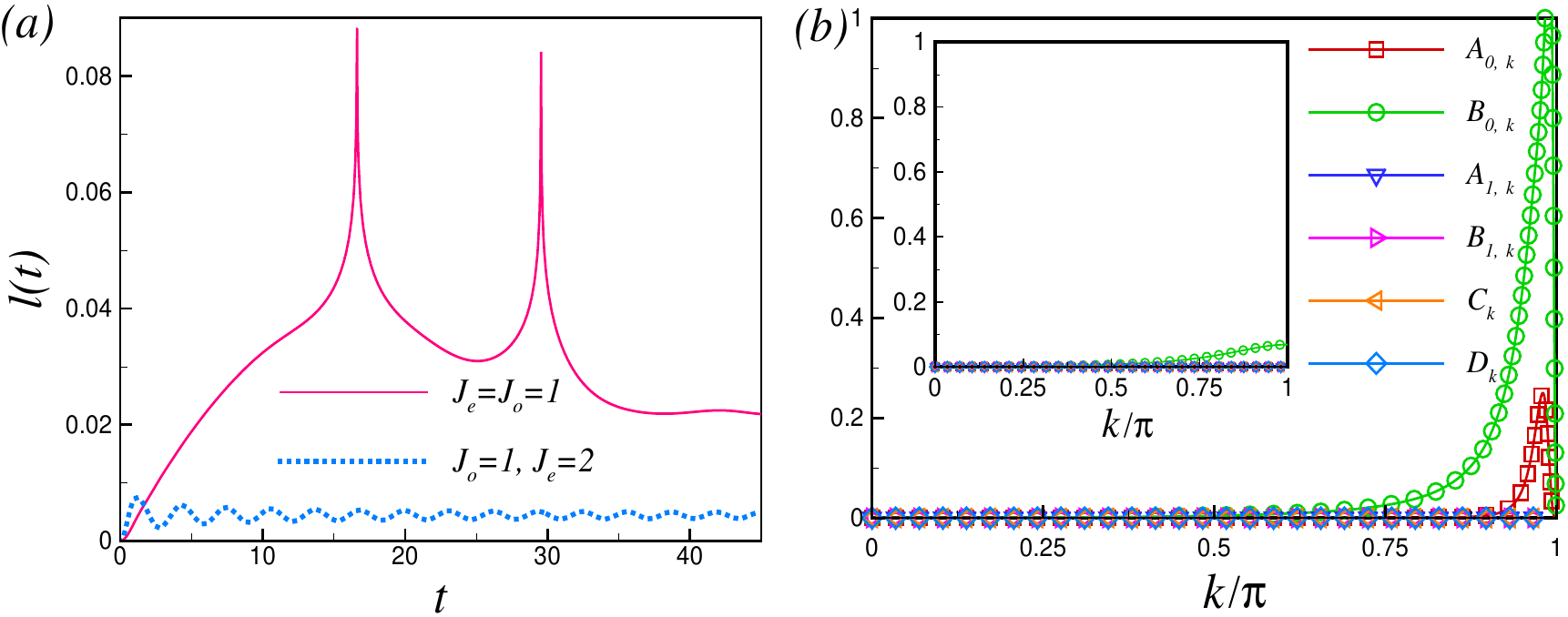}}
\caption{ (Color online) (a) The presence and the absence of DPTs
following a small size sudden quenching of $\theta$ in the rate function of return
probability. The solid line corresponds
to the IP where there are periodic occurrences of DPTs while DPTs get
rounded of away from the IP (dotted line).
(b) The amplitudes of oscillation in Eq. \ref{eq10},
for a small size quench across the critical point, plotted versus $k$ at the IP $J_{o}=J_{e}=1$.
Inset: The amplitudes of oscillation away from the isotropic point $J_{o}=1, J_{e}=2$.}
\label{fig4}
\end{figure}
%
%
\begin{figure}[ht]
\centerline{\includegraphics[width=\linewidth,height=0.38\linewidth]{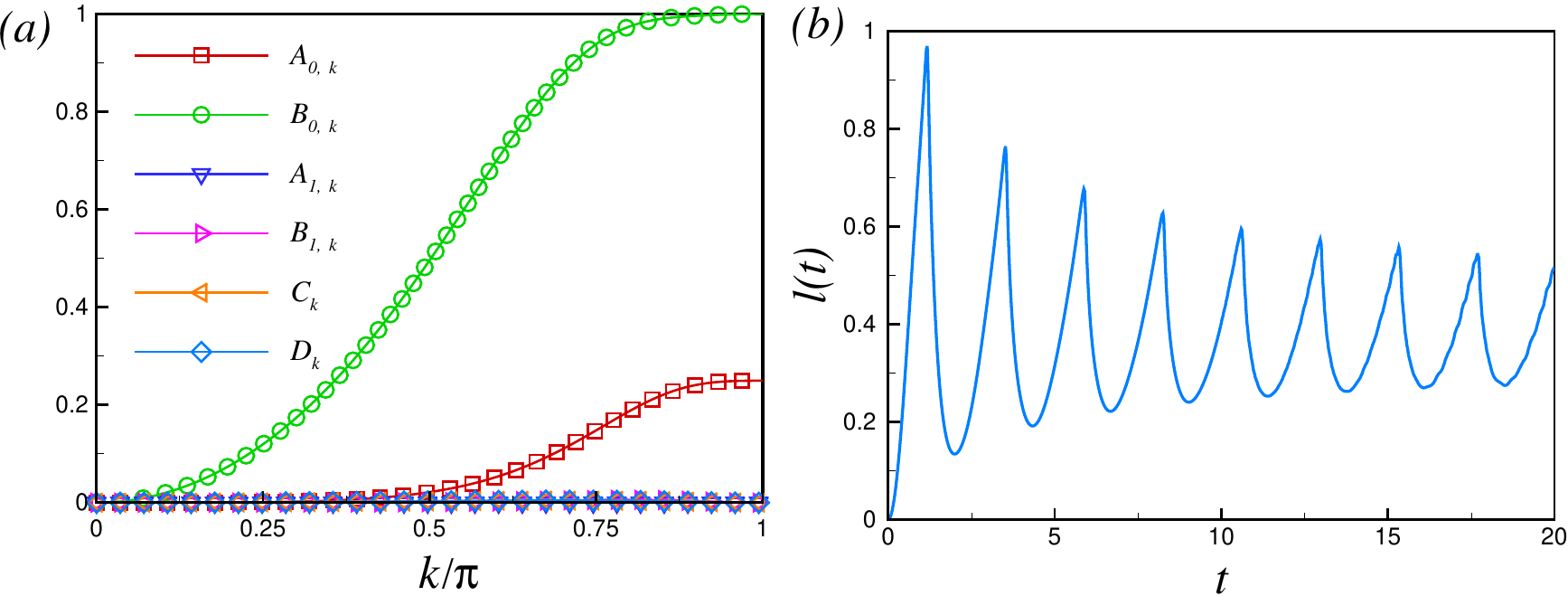}}
\caption{ (Color online) (a) The amplitudes of oscillation in Eq. \ref{eq10},
for a large size quench crossing the critical point, plotted versus $k$
away from the isotropic point $J_{o}=1, J_{e}=2$. (b) The rate function of the return
probability $l(t)$, for a
large size quench from $\theta_{1}=0.4\pi$ to $\theta_{2}=0.6\pi$ for system size
$N=120$, away from the IP $J_{o}=1, J_{e}=2$. Cusps in  $l(t)$ are clearly
visible indicating a dynamical phase transition.}
\label{fig5}
\end{figure}
%
By a rather lengthy calculation one can obtain the complete set of
eigenstates $|\psi_{m, k}(\theta)\rangle, (m=0,...,7)$ of the model
(for details, see the Appendix \ref{AppA}), yielding an exact expression for the LA and RP by sudden quench of
$\theta$ $(\theta_{1}\longrightarrow\theta_{2})$ \cite{JJ2017b,Jafari2016}
%
\begin{eqnarray}
\label{eq11}
Z(z)&=&\prod_{k>0}[\alpha_{3,k}+\alpha_{4,k}+\alpha_{5,k}+\alpha_{6,k}+\alpha_{1,k}e^{-z\varepsilon^{1}_{k}(\theta_{2})}
+\alpha_{2,k}e^{-z\varepsilon^{2}_{k}(\theta_{2})}+\alpha_{7,k}e^{z\varepsilon^{2}_{k}(\theta_{2})}+\alpha_{8,k}e^{z\varepsilon^{1}_{k}(\theta_{2})}].
\end{eqnarray}
%
%
\begin{eqnarray}
\label{eq12}
{\cal L}(t)&&=\prod_{k>0}
|1-A_{0,k}\sin^{2}[(\varepsilon_{k}^{1}(\theta_{2})+\varepsilon_{k}^{2}(\theta_{2}))t]
-B_{0,k}\sin^{2}[(\varepsilon_{k}^{1}(\theta_{2})+\varepsilon_{k}^{2}(\theta_{2}))t/2]
-A_{1,k}\sin^{2}[(\varepsilon_{k}^{1}(\theta_{2})-\varepsilon_{k}^{2}(\theta_{2}))t]\\
\nonumber
&&-B_{1,k}\sin^{2}[(\varepsilon_{k}^{1}(\theta_{2})-\varepsilon_{k}^{2}(\theta_{2}))t/2]
-C_{k}\sin^{2}[\varepsilon_{k}^{2}(\theta_{2})t]-D_{k}\sin^{2}[\varepsilon_{k}^{1}(\theta_{2})t]|,
\end{eqnarray}

where, $A_{0,k}, B_{0,k}, A_{1,k}, B_{1,k}, C_{k}$, and $D_{k}$ are function of overlaps between $k$ modes
of the initial ground state and eigenstates of the postquenched Hamiltonian
$\alpha_{m,k}=|\langle\psi_{m,k}(\theta_{2})|\psi_{0,k}(\theta_{1})\rangle|^{2}$ $(m=0,...,7)$ (for details, see the
Appendix \ref{AppB}).
%
%
The rate function of the RP following the quench from $\theta_{1}=0.49\pi$ to $\theta_{2}=0.51\pi$
is shown in Fig. \ref{fig4} (a) for the IP and away from the IP for system size $N=120$. Cusps in $l(t)$ are clearly visible as
an indicator of DPTs for the quench across the critical point $\theta_{c}=\pi/2$ at the IP while
nonanalyticities wiped out for the same quench away from the IP which reflects no DPT.
As seen in Eq. (\ref{eq11}), the LA is not a simple function of $z$ variable and then we can not obtain the zeros of LA analytically.
So, to obtain the real time nonanalyticities in the rate function of RP we have to investigate Eq. (\ref{eq12}) directly.
As discussed, the nonanalyticities in the rate function of the RP occur when the oscillation
amplitude, in the mode decomposition of the RP in Eq. (\ref{eq12}), takes its maximum possible value.
The oscillation amplitudes have been plotted in Fig. \ref{fig4}(b) for a small size quench from
$\theta_{1}=0.49\pi$ to $\theta_{2}=0.51\pi$ for both the IP ($J_{o}=J_{e}=1$) and away from the IP ($J_{o}=1, J_{e}=2$).
As seen, $A_{0,k}$ and $B_{0,k}$ is nonzero at the IP and $A_{0,k}$ takes its maximum possible value
at $k^{\ast}$ mode ($A_{0,k^{\ast}}=1$), while the oscillations amplitude are zero or very small away from
the IP (inset, Fig. \ref{fig4}(b)). Therefore, there is no mode where its contribution becomes zero in the product in
Eq. (\ref{eq12}) for resulting nonanalyticities in the $l(t)$ away from the IP. As oscillation amplitudes
are function of the probability of quasiparticle excitation, the different behaviours of $l(t)$, for a small size quench
across the critical point, at the IP and away from the IP, originates from the difference between excited states occupancies at
the IP and away from the IP.

As mentioned, the energy gap between the $\varepsilon^{2}_{k}$ and $\varepsilon^{3}_{k}$ bands,
which control the phase transition, is closed for all $k$ mode at $\theta_{c}=\pi/2$ for arbitrary values of $J_{e}/J_{o}$.
While the energy gap between the $\varepsilon^{1}_{k}$ and $\varepsilon^{4}_{k}=-\varepsilon^{1}_{k}$ bands closes at $k=\pi$, $\theta_{c}=\pi/2$
and it is nonzero away from the IP.

Since, a sudden quench generally leads to particle-hole type excitations, it is expected that
a small quenches, which puts a small amount of energy into the system, yields a large oscillation amplitude at the
IP due to dispersionless quasiparticle band at $k=\pi$, $\theta_{c}=\pi/2$. However, we expect a small oscillation amplitude away from
the IP due to gapped quasiparticle band at $k=\pi$, $\theta_{c}=\pi/2$.
The contribution of the $k^{\ast}$ mode in Eq. (\ref{eq12}) becomes zero at time instances at which $A_{0,k^{\ast}}\sin^{2}[(\varepsilon^{1}_{k^{\ast}}(\theta_{2}))t]=1$.
Then, the real time nonanalyticities for a small quench across the critical line is given by $t_{n}=t^{\ast}(n+\frac{1}{2})$, where $t^{\ast}\simeq\pi/\varepsilon^{1}_{k^{\ast}}$.
It should be noted that, the oscillation function correspond to $B_{0,k^{\ast}}$  ($\sin^{2}[(\varepsilon^{1}_{k^{\ast}}(\theta_{2}))t/2]$) is zero at time $t_{n}$.

However, for a large size quench which imposes a large amount of energy into the system, and in turn increases the
probability of quasiparticle excitation, we expect a large oscillation amplitudes and consequently nonanalyticities
in $l(t)$ even away from the IP where the quasiparticle energy is gapfull.
The oscillations amplitude have been plotted in Fig. \ref{fig5}(a) for a large size quench from
$\theta_{1}=0.4\pi$ to $\theta_{2}=0.6\pi$ away from the IP ($J_{o}=1, J_{e}=2$).
As expected, the oscillation amplitude $A_{0,k}$ reaches its maximum possible value at $k^{\ast}$ which results nonanalyticities
in $l(t)$ (Fig. \ref{fig5}(b)).
The real time nonanalyticities for a large quench crossing the critical line is given by $t_{n}=t^{\ast}(n+\frac{1}{2})$, where $t^{\ast}=\pi/(\varepsilon^{1}_{k^{\ast}}+\varepsilon^{2}_{k^{\ast}})$.

%
%
We should stress that, the most pronounced revivals in the RP happen when the system satisfies two circumstances,
large oscillation amplitude (maximum possible value is not necessary) and the zero energy mode \cite{JJ2017b, Zhanga, Zhangb}, while occurrence
of the DPTs only needs large oscillation amplitude with maximum possible value $1$.
%

\section*{Summary and conclusions}
We have shown that the presence of quantum phase transition point is neither a sufficient nor a necessary condition for observing a dynamical quantum phase transition after a global quantum quench.
By examining how the eigenstates of the models imprint the return probability, we find that what does matter is the availability of propagating quasiparticles as signaled by their having
an impact on the rate function of the return probability. Searching the dynamical phase transition in the extended XY model, provides an example that a stable massless phase can act as a source of dynamical phase transition.
While a quantum phase transition generically supports massless excitations, our case study of the extended quantum compass model reveals that these excitations may not necessarily couple to the quantum phase transition.

We should point out that, in Ref. \cite{Andraschko} it has been reported that in a transfer
matrix approach, nonanalyticities in rate function of the return probability are a consequence of crossing of the leading eigenvalue with the next
leading eigenvalue of the Hamiltonian for a quench within the same phase.
However, it also shown that, for a quench across the quantum phase transition point, any quench starting in the ferromagnetic phase and any quench where only the uniform magnetic field is changed, leads to zero rate function of the return probability \cite{Andraschko}.
The zero values of rate function of the return probability in the former case originates from the fact that the ferromagnetic state is an
eigenstate of both the pre-quenched and the post-quenched Hamiltonians \cite{Andraschko, Heylrev}.In the latter case the conservation of the total magnetization results zero rate function of the return probability.
In this paper the quench has not been done by changing the magnetic field and the initial state in both the extended $XY$ model and the extended compass model is not the eigenstate of the post-quenched Hamiltonian \cite{Andraschko, Heylrev}.
So, our findings may call for a revisit of earlier studies on dynamical phase transition and quantum criticality,
and can shed new light on the bridge between dynamical phase transition and quantum phase transitions.

\section*{Acknowledgements}
The author would like to thank Henrik Johannesson, Alireza Akbari and Utkarsh Mishra for reading the manuscript and
valuable comments.

\section{Appendix}

\subsection{General Compass model\label{AppA}}

The EQCC ground state $|\psi_0\rangle$ is realized by filling up the negative-energy quasiparticle states, $|\psi_0\rangle = \prod_k \gamma_k^{(1) \dag} \gamma_k^{(2) \dag} |0\rangle$, where $|0\rangle$ is the Bogoliubov vacuum annihilated by the $\gamma_k$:s \cite{Jafari2016}. While excited states can be similarly obtained, their construction becomes quite cumbersome within the Bogoliubov-de Gennes formalism. An alternative approach was pioneered by Sun \cite{Sun2009}. One here takes off from the observation that the QCC Hamiltonian can be written as a sum of commuting Hamiltonians $H_k$,
\begin{eqnarray}
\label{commH}
H_k=J_{k}c_{k}^{A\dag}c_{-k}^{B\dag}+L_{k}c_{k}^{A\dag}c_{k}^{B} +J_{-k}c_{-k}^{A\dag}c_{k}^{B\dag}+L_{-k} c_{-k}^{A\dag}c_{-k}^{B}+\mbox{H.c},
\end{eqnarray}
Since $H_k$ conserves the number parity (even or odd number of electrons), it is sufficient to consider the even-parity
subspace of the Hilbert space, spanned by
\begin{align}
\nonumber
|\varphi_{1,k}\rangle&=|0\rangle,\!&\!|\varphi_{2,k}\rangle&=c_{k}^{A\dag}c_{-k}^{A\dag}|0\rangle,~|\varphi_{3,k}\rangle=c_{k}^{A\dag}c_{-k}^{B\dag}|0\rangle,\\
\nonumber
|\varphi_{4,k}\rangle&=c_{-k}^{A\dag}c_{k}^{B\dag}|0\rangle,~\!&\!|\varphi_{5,k}\rangle&=c_{k}^{A\dag}c_{-k}^{B\dag}|0\rangle, ~|\varphi_{6,k}\rangle=c_{k}^{A\dag}c_{k}^{B\dag}|0\rangle,\\
\label{eq7}
|\varphi_{7,k}\rangle&=c_{-k}^{A\dag}c_{-k}^{B\dag}|0\rangle,~\!&\!|\varphi_{8,k}\rangle&=c_{k}^{A\dag}c_{-k}^{A\dag}c_{k}^{B\dag}c_{-k}^{B\dag}|0\rangle.
\end{align}
Given this basis, the eigenstates $|\psi_{m,k}\rangle$ of $H_k$ can be written as $|\psi_{m,k}\rangle=\sum_{j=1}^{8}v_{m,k}^{(j)}|\varphi_{j,k}\rangle,$

\subsection{Loschmidt echo\label{AppB}}
The amplitudes in the mode decomposition of the RP, Eq. (12), depend on the state overlaps $\alpha_{m,k}=|\langle\psi_{m,k}(\theta_{2})|\psi_{0,k}(\theta_{1})\rangle|^{2}$ $(m=0,...,7)$ as

\begin{eqnarray}
\nonumber
A_{0,k}&=&4\alpha_{0,k}\alpha_{7,k},\\
\nonumber
B_{0,k}&=&4(\alpha_{2,k}+\alpha_{3,k}+\alpha_{4,k}+\alpha_{5,k})(\alpha_{0,k}+\alpha_{7,k}),\\
\nonumber
A_{1,k}&=&4\alpha_{1,k}\alpha_{6,k},\\
\nonumber
B_{1,k}&=&4(\alpha_{2,k}+\alpha_{3,k}+\alpha_{4,k}+\alpha_{5,k})(\alpha_{1,k}+\alpha_{6,k}),\\
\nonumber
C_{k}&=&4(\alpha_{0,k}\alpha_{1,k}+\alpha_{6,k}\alpha_{7,k}),\\
\nonumber
D_{k}&=&4(\alpha_{0,k}\alpha_{6,k}+\alpha_{1,k}\alpha_{7,k}).
\end{eqnarray}


\section*{Additional information}

\textbf{Contributions}: All parts of this paper have been done by R. Jafari.

\textbf{Competing interest}: The author declare no competing interests.

\end{document}